\documentclass[conference]{IEEEtran}
\IEEEoverridecommandlockouts

\usepackage{cite}
\usepackage{amsmath,amssymb,amsfonts}
\usepackage{graphicx}
\usepackage{textcomp}
\usepackage{booktabs}
\usepackage{url}
\usepackage{enumitem}

\graphicspath{{figures/}}

\begin{document}

\title{Multistatic Terahertz ISAC for Object Shape Identification \\
\thanks{This work was funded by the NSF under award CNS-2416077.}}

\author{\IEEEauthorblockN{Ralph Kumah Assan}
\IEEEauthorblockA{\textit{Dept.\ of Electrical and Computer Engineering}\\
\textit{Portland State University}\\ Portland, OR 97207, USA\\ rassan@pdx.edu}
\and
\IEEEauthorblockN{Suresh Singh}
\IEEEauthorblockA{\textit{Dept.\ of Computer Science}\\
\textit{Portland State University}\\ Portland, OR 97207, USA\\ singh@cs.pdx.edu}}

\maketitle

\begin{abstract}
6G networks are expected to sense their surroundings using the same waveforms that carry data. We ask how much can be learned about the shape of an object from the terahertz echoes that user equipment transmissions produce as they scatter from that object on their way to a ceiling-mounted access point, with no sensing hardware beyond the communication link itself. We build a physics-based multistatic forward model at $f_c=300$\,GHz with $50$\,GHz of bandwidth, combining Physical Optics (PO) and Physical Theory of Diffraction (PTD) scattering with an image-source multipath lattice, molecular absorption and directional horn antennas. The model makes explicit a property of the PO mechanism hierarchy that is usually left implicit. Six physically distinct mechanisms share only three frequency exponents, so the spectral slope does not by itself determine which mechanisms a target presents and therefore cannot by itself determine its geometry. Angular diversity is accordingly a requirement for identifiability rather than an optional refinement. We compare a transmitter-count-invariant 2D-CNN with a physics-informed RBF-SVM on object orientations never seen in training. Accuracy climbs from roughly $70\%$ with one transmitter to $96$--$97\%$ with $20$. The CNN gains in two regimes, irregular placement and lossy targets, while on metal under quasi-uniform placement the SVM baseline is fully competitive. Performance falls off gradually under occlusion, holding $78\%$ with three quarters of the surface hidden and $63\%$ when the classifier has never seen an occluded target. Multi-object scenes are limited by interference between overlapping returns rather than by angular coverage, so presence detection saturates after roughly four transmitters.
\end{abstract}


\section{Introduction}

The sixth generation (6G) of wireless networks is expected to carry data at very high rates while at the same time sensing the physical environment \cite{liu2022isac,chen2021thz,itu2160}. This dual role is formalized as Integrated Sensing and Communication (ISAC), which the ITU designates as one of six core usage scenarios for 6G \cite{itu2160} and which 3GPP has studied since Release~19 \cite{3gpp22837}. One prominent application is the smart environment, where access points model the space around them to support digital twins, activity recognition and robot navigation without dedicated sensors \cite{gong2025,etsi2025}. Terahertz bands suit this role well because they offer both large bandwidth and short wavelength \cite{liu2022limits,han2024thzisac}.

Most terahertz sensing work pursues reconstruction, recovering a spatially resolved image from which geometry and material can then be read off \cite{li2023,sheen2001,batra2022,heimbeck2014,lyu2024}. That is the right tool when the target is unknown, but many practical tasks ask a narrower question. When the set of objects that can appear is known in advance, the system needs only to decide which one it is looking at. A warehouse holds cuboidal packages, cylindrical drums and conical hazard markers, and a logistics robot needs the class of the object rather than its image in order to choose an action. Learned classifiers have addressed this question for radar and terahertz targets before, but almost always from imaged or range-profile inputs \cite{batra2022}. In this work the observation is instead an opportunistic multistatic echo of ordinary communication traffic.

We develop a multistatic THz-ISAC framework that classifies the shape and material of an object from the waveforms that spatially distributed user equipments (UEs) already transmit. Our specific contributions are as follows.

\begin{itemize}[leftmargin=*,itemsep=1pt,topsep=2pt,parsep=0pt]
\item We develop a physics-based multistatic terahertz forward model and make explicit that the spectral-hierarchy argument for shape discriminability is \emph{degenerate}. Mechanisms sharing an exponent make the spectral slope necessary but not sufficient, which reframes angular diversity as a condition for identifiability.

\item We establish an evaluation protocol in which training and test partitions occupy \emph{disjoint object-pose bins}, so reported accuracy measures generalization to unobserved orientations.

\item We compare the CNN against a physics-informed RBF-SVM and identify the two regimes in which the learned model wins clearly, namely irregular transmitter placement and lossy materials. Under quasi-uniform placement on metal the classical baseline matches it.

\item We extend the framework to multi-object scenes, showing that they are \emph{interference-limited} rather than diversity-limited, and we characterize occlusion for classifiers that have and have not seen blockage during training.
\end{itemize}

\section{System Model}

\begin{figure}[t]
\centering
\includegraphics[width=0.78\columnwidth]{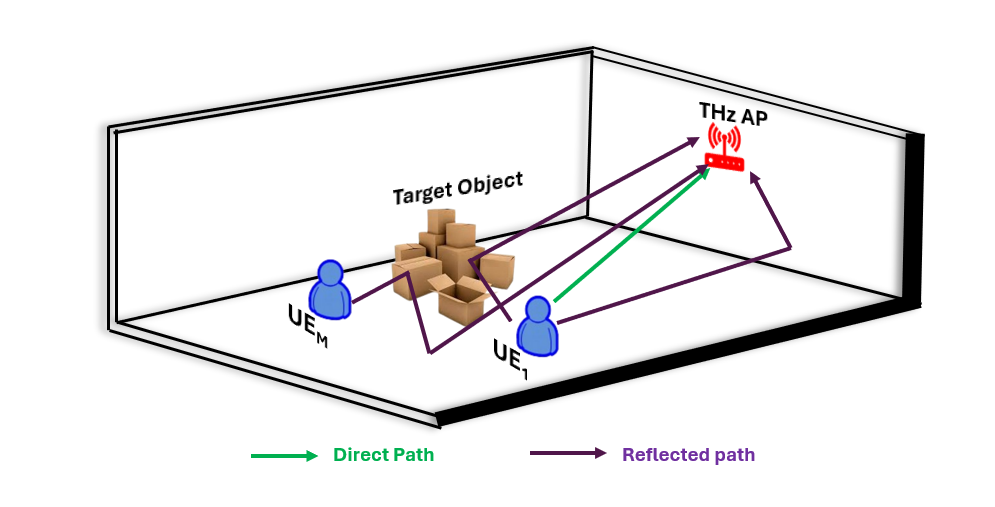}
\caption{Proposed THz-ISAC multistatic shape identification system. UEs at
different locations transmit to the ceiling-mounted AP, which receives both the
direct path used for data decoding and the reflected path via the target used
for sensing. Each UE provides a distinct bistatic vantage point.}
\label{fig:system}
\end{figure}

The architecture is illustrated in Fig.~\ref{fig:system}. UEs distributed across an indoor environment transmit towards a ceiling-mounted access point (AP), and each transmission propagates along a direct path, a reflected path via the target, or both. The reflected component is not a separate sensing transmission but an inherent by-product of the communication signal, and it encodes shape-dependent information in its amplitude, phase and frequency structure. Reflections from $M$ spatially diverse transmitters together form a multistatic observation whose joint spectral-angular structure is characteristic of the object geometry.

\subsection{Room Geometry and Antenna Configuration}

The environment is a $[6,8,3]$\,m rectangular room. The AP is at $\mathbf{r}_{\mathrm{AP}}=[3,4,2.95]$\,m and the target at $\mathbf{r}_{\mathrm{obj}}=[3,4,0.75]$\,m with nominal height $h=0.50$\,m. Per-sample diversity comes from azimuthal rotation $\phi\in[0^\circ,360^\circ)$, height scaling $\eta\sim\mathcal{U}[0.90,1.10]$ applied to the object \emph{geometry}, and position jitter $\delta\mathbf{r}\sim\mathcal{N}(\mathbf{0},(0.05)^2\mathbf{I}_3)$\,m. Scaling the geometry rather than the returned amplitude is essential, since a $10\%$ change in a $0.50$\,m object displaces scattering centers by up to $5$\,cm, which at $\lambda=1$\,mm is $50$ wavelengths of one-way path. Object size therefore acts predominantly as a phase effect and cannot be represented as an amplitude rescaling. Both TX and AP antennas are aperture-type horns with a sinc-squared pattern,
\begin{equation}
G(\vartheta)=\left[\frac{\sin u}{u}\right]^{2},\qquad
u=\frac{\pi D}{\lambda}\sin\vartheta ,
\label{eq:horn}
\end{equation}
where $\vartheta$ is the off-boresight angle and $D/\lambda=1.391/\!\left(\pi\sin(\theta_{\mathrm{3dB}}/2)\right)$ with $\theta_{\mathrm{3dB}}=15^\circ$. A $-20$\,dB sidelobe floor is imposed beyond the first null, back-hemisphere radiation is suppressed by $\cos^{2}\vartheta$, and both boresights are steered toward the object. We emphasize that \eqref{eq:horn} is evaluated against the antenna boresight and not against the local surface normal.

Two placement strategies are compared. Random placement draws UE positions uniformly within the room, while deterministic Fibonacci-sphere placement \cite{roberts2018} yields quasi-uniform angular coverage. Both sample the \emph{same} admissible region, which is the room interior with a $0.25$\,m wall margin, a minimum object distance of $1.2$\,m and a UE height band of $[0.25,2.60]$\,m. Fibonacci directions are realized by ray-box intersection so each direction is preserved exactly, and directions that cannot satisfy the minimum distance are discarded before the set is thinned to $M$ points by greedy angular separation. Under both strategies the constellation is redrawn independently for every sample, so no fixed room realization can be memorized.

\subsection{Forward Channel Model}
\label{sec:channel}

The probing signal spans $B=50$\,GHz at $f_c=300$\,GHz with $N_f=64$ bins, giving range resolution $\Delta r\approx3$\,mm. The received signal at bin $n$ from transmitter $m$ is
\begin{align}
Y(m,n)=\!\!\sum_{\ell\in\mathcal{L}_{\mathrm{vis}}}\!\!
&\Gamma^{p_\ell}\,\rho^{\,q_\ell/2}\, G^{\mathrm{TX}}_\ell G^{\mathrm{RX}}_\ell
S_\ell(f_n)\,L_{\mathrm{atm}}(f_n,d_\ell)\nonumber\\
&\times\cos\theta^{\mathrm{i}}_\ell\cos\theta^{\mathrm{s}}_\ell\,
\frac{e^{-j2\pi f_n d_\ell/c}}{d^{\mathrm{TX}}_\ell d^{\mathrm{AP}}_\ell}
+N(m,n),
\label{eq:channel}
\end{align}
where $\mathcal{L}_\text{vis}$ is the set of visible surface patches, $\Gamma_\ell$ the Fresnel reflection coefficient and $p_\ell\in\{1,2,3\}$ the number of target interactions along that path, $\rho$ the wall power reflectivity and $q_\ell$ the number of wall bounces, $G_\ell^\text{TX}$ and $G_\ell^\text{RX}$ the horn gains, $S_\ell(f_n)$ the PO scattering amplitude, $L_\text{atm}=10^{-\alpha(f_n)d_\ell/20}$ the atmospheric transmission factor, $\cos\theta^{\mathrm{i}}_\ell\cos\theta^{\mathrm{s}}_\ell$ the projection of patch area onto the incident and scattered directions, and $N(m,n)\sim\mathcal{CN}(0,\sigma_n^2)$ additive complex Gaussian noise at 20\,dB SNR. The total path length is $d_\ell=d^{\mathrm{TX}}_\ell+d^{\mathrm{AP}}_\ell$, so the spreading term uses the product of the two legs while absorption and phase use their sum. Equation~\eqref{eq:channel} models the target-scattered component only.

\subsection{Propagation Environment and Materials}

Wall and floor reflections use the image-source method, in which each wall bounce is replaced by a mirrored transmitter copy. Reflections up to second order are retained, and a path that has bounced $p$ times is attenuated by $\rho^{p/2}$ with $\rho = 0.45$, the power reflectivity of dry plaster. Atmospheric absorption follows a sum of Lorentzian line shapes at the principal water-vapor resonances, calibrated to ITU-R P.676 \cite{itup676}, which gives roughly $33$\,dB/km at $300$\,GHz. The upper band edge approaches the $325$\,GHz absorption line and imposes a spectral tilt that is the same for every shape, and the per-sample normalization of Section~\ref{sec:input} removes it. Material enters through $\Gamma=(\sqrt{\varepsilon_c}-1)/(\sqrt{\varepsilon_c}+1)$ with $\varepsilon_c=\varepsilon_r(1-j\tan\delta)$, which gives $|\Gamma|=1$ for metal, $|\Gamma|=0.226$ for HDPE plastic and $|\Gamma|=0.190$ with $\tan\delta=0.150$ for lossy wood.

\begin{table}[t]
\caption{PO/PTD mechanism amplitudes and frequency exponents}
\label{tab:mech}
\centering
\scriptsize
\setlength{\tabcolsep}{3pt}
\begin{tabular}{@{}lll@{}}
\toprule
Mechanism & Amplitude $S$ & $|S|\propto$ \\
\midrule
Flat facet & $\dfrac{jkA}{2\pi}\operatorname{sinc}\dfrac{ku_1L_1}{2}\operatorname{sinc}\dfrac{ku_2L_2}{2}$ & $k$ \\[6pt]
Singly curved & $\dfrac{j}{2\pi}L\sqrt{2\pi\rho_ck/w}\;e^{-j\pi/4}$ & $k^{1/2}\sqrt{\rho_c}$ \\[6pt]
Doubly curved & $\sqrt{\rho_1\rho_2}/2$ & $k^{0}$ \\[2pt]
Edge (PTD)    & $D(\beta,\beta')L_e\operatorname{sinc}\dfrac{ku_eL_e}{2}\,e^{-j\pi/4}$ & $k^{0}$ \\[6pt]
Trihedral     & $C_t\,k\,a^{2}\,g(\theta_{\mathrm{r}},\theta_{\mathrm{a}})$ & $k$ \\[2pt]
Tip / apex    & $C_p\sqrt{A_{\mathrm{tip}}}$ & $k^{0}$ \\
\bottomrule
\end{tabular}
\vspace{1pt}

\scriptsize At specular incidence the facet sinc factors approach unity and
$\sigma=4\pi A^{2}/\lambda^{2}$, the standard plate result \cite{knott2004},
while the doubly curved entry gives $\sigma=\pi a^{2}$ for a sphere. Six
mechanisms occupy only three exponent values. The edge exponent is $k^{0}$
because $D$ carries no frequency dependence, so its sinc envelope varies the
contribution with angle rather than frequency, and that angular selectivity is
precisely what a spectral slope discards.
\end{table}

\section{Physical Optics Scattering Model}
\label{sec:po}

Each surface element is treated in the Physical Optics approximation \cite{felsen1994}. Every mechanism is written as a scattering amplitude $S$ with units of length, related to radar cross section by $\sigma=4\pi|S|^{2}$, which lets each frequency exponent be checked by dimensional analysis. Throughout, $k=2\pi f/c$ and $\hat{\mathbf{u}}=\hat{\mathbf{s}}-\hat{\mathbf{i}}$ is the bistatic difference vector with components $u_1,u_2,u_e$ along the relevant extents, and $A$ is element area. Surfaces are discretized into patches small enough that the PO integrand is locally stationary across each one, giving $152$ patches for the cube, $205$ for the cone and $264$ for the cylinder.

Table~\ref{tab:mech} collects the six mechanisms. Three entries need comment. For the singly curved surface, $w=|\hat{\mathbf{u}}\cdot\hat{\mathbf{c}}|$ measures how far the observation direction lies from broadside along the curved coordinate $\hat{\mathbf{c}}$, and since the stationary-phase expansion breaks down as $w\to0$ we floor $w$ at a small value chosen to reproduce the broadside limit $\sigma=2\pi\rho_cL^{2}/\lambda$. No target in our library is doubly curved, but that mechanism completes the taxonomy and will be needed once the vocabulary extends beyond convex primitives. For edge diffraction, $\beta$ and $\beta'$ are the incidence and diffraction angles measured from the edge, distinct from the object azimuth, and the Ufimtsev fringe coefficient $D$ \cite{ufimtsev2007} carries no frequency dependence, which fixes the $k^{0}$ scaling while the sinc factor confines the contribution to the Keller cone \cite{keller1962}.

The trihedral corner needs a gate. A trihedral reflects strongly only within roughly $\pm30^\circ$ of its axis, a condition rarely met in a floor-to-ceiling geometry, so applying its full backscatter strength at arbitrary angles would overstate the return by orders of magnitude. The Gaussian gate $g=\exp(-\theta_{\mathrm{r}}^{2}/2\sigma_{\mathrm{r}}^{2})\exp(-\theta_{\mathrm{a}}^{2}/2\sigma_{\mathrm{a}}^{2})$ suppresses the contribution outside that window, where $\theta_{\mathrm{r}}$ is the angle between $\hat{\mathbf{s}}$ and $-\hat{\mathbf{i}}$ and $\theta_{\mathrm{a}}$ the angle between the trihedral axis and $-\hat{\mathbf{i}}$. The widths $\sigma_{\mathrm{r}}=25^\circ$ and $\sigma_{\mathrm{a}}=30^\circ$ match the half-power extent reported in \cite{knott2004} and were not tuned to the classification results. The gating does not matter for the cube, which stays well separated even with the corner term suppressed, because what distinguishes it is the angular sparsity of its specular lobes rather than its retro-reflection. The constants $C_t$ and $C_p$ are fixed once so that the trihedral attains the canonical maximum $\sigma=12\pi a^{4}/\lambda^{2}$ \cite{knott2004} on axis.

\subsection{Aggregate Spectral Model and Mechanism Degeneracy}
\label{sec:degeneracy}

Summing the mechanisms, the noise-free spectrum of shape $s$ is
\begin{equation}
\bar{Y}_s(n)=a^{(0)}_s+a^{(1/2)}_s k_n^{1/2}+a^{(1)}_s k_n ,
\label{eq:aggregate}
\end{equation}
where $a^{(\beta)}_s$ collects every mechanism with exponent $\beta$. The logarithm of the dominant term, $\log|\bar{Y}_s|\approx\beta_s\log k_n+\log a^{(\beta)}_s$, is a straight line in log-log space whose gradient $\beta_s$ is estimated by least squares. Fitting $\beta_s$ is the natural way to ask what a target is made of.

The third column of Table~\ref{tab:mech} shows why that fit cannot answer the question on its own, since six mechanisms occupy only three exponent values. The scaling of each individual mechanism is classical \cite{knott2004,ufimtsev2007,keller1962} and what we draw out is its consequence for classification. Two objects built from entirely different mechanisms can return the same slope, so estimating $\beta_s$ narrows the candidate set without ever resolving it, and no amount of bandwidth or SNR changes that.

\begin{figure}[t]
\centering
\includegraphics[width=0.74\columnwidth]{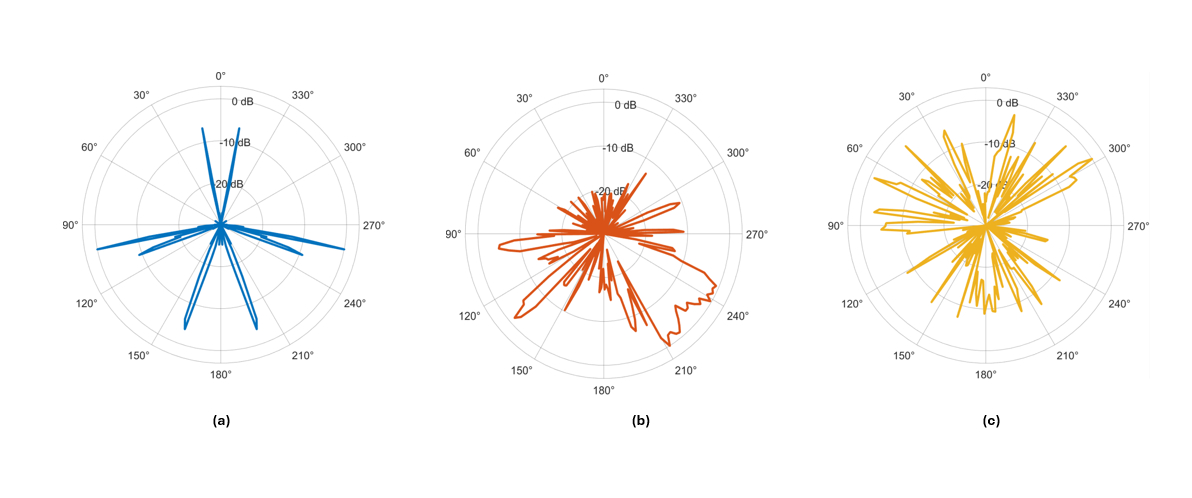}
\caption{Bistatic scattering patterns in the horizontal plane at $300$\,GHz for a metal PEC target. (a)~Cube, showing narrow specular lobes at face normals with deep nulls between them. (b)~Cone, showing nearly isotropic radiation from the apex with a diffuse base ring. (c)~Cylinder, showing a broad broadside lobe with two sharp rim-diffraction peaks. The radial axis is normalized $|S|$ in dB over a $40$\,dB span.}
\label{fig:bistatic}
\end{figure}

What separates mechanisms sharing an exponent is where they send their energy. A flat facet radiates into a narrow specular lobe, a trihedral into a retro-directed cone, an edge onto the Keller cone, and a doubly curved surface almost isotropically. Fig.~\ref{fig:bistatic} shows the consequence for whole objects, where cube, cone and cylinder produce clearly distinct bistatic patterns despite overlapping spectral slopes. A single transmitter samples one direction and cannot see this structure, whereas a multistatic constellation samples many and can. Angular diversity is therefore a requirement for identifiability rather than a refinement of accuracy, and the sections that follow test this claim empirically.

\section{CNN Classifier vs RBF-SVM Baseline}

\subsection{Input Representation and Architecture}
\label{sec:input}

Each sample is a real-valued tensor $\mathbf{X}_i\in\mathbb{R}^{N_f\times M_{\max}\times2}$ with $M_{\max}=20$. The frequency axis captures the PO spectral structure, each transmitter-axis column adds an independent bistatic viewpoint, and the two channels retain real and imaginary parts. Per-sample RMS normalization removes material-dependent signal strength so the network learns shape signatures rather than absolute magnitude, and a separate model is trained per $M$ using only the first $M$ columns, which ensures that gains with $M$ reflect angular information rather than a larger training set.

The 2D-CNN uses three Conv2D$\rightarrow$BatchNorm$\rightarrow$ReLU blocks with $32$, $64$ and $64$ filters and kernel $[5\times f_t]$, $f_t=\min(3,M)$. A MaxPool$[2\times1]$ layer halves frequency resolution while preserving the transmitter axis, and Global Average Pooling collapses the feature volume to a fixed $\mathbb{R}^{64}$ vector regardless of $M$, making the architecture transmitter-count invariant with roughly $87{,}000$ parameters at $M=12$. Columns are ordered by room position, so the network could in principle weight column indices differently. We state this as a hypothesis rather than an established mechanism, since UE coordinates are not an input and a column index carries only an order statistic once the constellation is redrawn per sample. All results are averaged over three independent seeds with the best-validation-loss model retained, and shaded bands denote one standard deviation.

\begin{figure}[t]
\centering
\includegraphics[width=0.74\columnwidth]{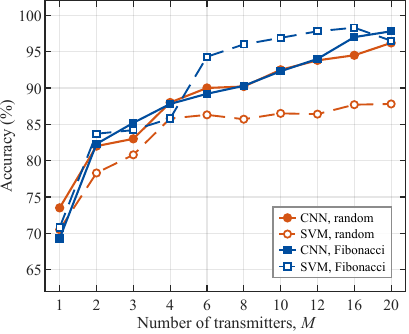}
\caption{Overall shape identification accuracy versus transmitter count $M$ at $300$\,GHz, metal target, $20$\,dB SNR, with disjoint train and test azimuth bins. Solid lines show the CNN and dashed lines the RBF-SVM. Random-placement curves are averaged over three room realizations and three seeds each.}
\label{fig:A1}
\end{figure}

\subsection{Baseline Classifier}
\label{sec:baseline}

To assess whether a learned model is necessary we build an RBF-SVM baseline from physics-derived features. These are the log-log slope, intercept and curvature of \eqref{eq:aggregate}, the spectral centroid, spectral flatness, the upper-to-lower band energy ratio, the lag-1 autocorrelation of $|Y|$, and the peak-to-mean ratio of the range profile $|\mathrm{IFFT}(Y)|$. They are aggregated across the $M$ links by mean, standard deviation, minimum, maximum and median, which makes the feature vector transmitter-count invariant in the same way global average pooling is.

Two asymmetries separate the classifiers and they are not of the same kind. First, the SVM aggregation is permutation invariant across transmitters and so discards which UE produced which measurement, whereas the CNN retains that ordering. Second, and more consequentially, every SVM feature derives from $|Y|$ or $|\mathrm{IFFT}(Y)|$, whereas the CNN receives real and imaginary parts and therefore retains relative phase. The comparison measures representation together with input richness rather than representational capacity alone, and we return to this when interpreting the material results.

\section{Results}

All results are at $f_c=300$\,GHz and $20$\,dB per-bin SNR with disjoint pose partitions, using $27$ training and $9$ test orientations.

\subsection{Transmitter Count, Placement and Classifier Comparison}
\label{sec:resA}

Fig.~\ref{fig:A1} presents overall accuracy against transmitter count, and the first result is that angular diversity delivers real information gain. All four configurations rise from $69$--$74\%$ at $M=1$ to $88$--$97\%$ at $M=20$, with the steepest gain over $M\in[1,4]$ and little improvement beyond $M\approx8$. Because a separate model is trained per $M$ on a fixed dataset this gain cannot be an artifact of training set size, and because train and test poses are disjoint it cannot be memorization of orientation.

The two placement strategies separate the classifiers. Under Fibonacci placement the SVM is fully competitive, reaching $97.7\%$ at $M=12$ against the CNN's $93.8\%$, with the two within a point of each other at $M=20$. Quasi-uniform placement makes the transmitters approximately exchangeable, so no single UE occupies a privileged angular position and the permutation-invariant aggregation of the SVM is close to sufficient. Such placement however requires controlled positioning that is not achievable in a real deployment. Under random placement some UEs fall near face-normal specular lobes while others land in deep nulls, so transmitter identity now carries substantial discriminative information. The SVM discards it by design and plateaus near $87\%$ from $M\approx6$ onward, while the CNN continues to improve and reaches $96.0\%$ at $M=20$, a gap of $8.3$ points. Since UE positions are uncontrolled in practice, random placement is the more realistic assumption, though we stop short of a deployment recommendation given the single forward model and the input asymmetry of Section~\ref{sec:baseline}.

Breaking the results down by shape shows where the difficulty lies. The cube separates almost perfectly from a small number of transmitters, while the cone and cylinder converge more slowly and account for essentially the entire residual error. A Bhattacharyya analysis \cite{bhattacharyya1943} makes this concrete, since the cube-to-cone and cube-to-cylinder distances are $0.549$ and $0.574$ whereas the cone-to-cylinder distance is only $0.034$, falling to $0.007$ in the hardest random realization. The cube is distinguished by the angular sparsity of its scattering pattern, a few intense lobes at face-normal directions separated by deep nulls, and that signature survives any placement strategy. The cone and cylinder are both bodies of revolution whose dominant returns differ only in the relative weight of a tip mechanism against a singly curved surface mechanism, rather than in which mechanism classes are present at all, so separating them requires accumulating several bistatic views.

\subsection{Per-Material Classification Accuracy}
\label{sec:resB}

\begin{figure}[t]
\centering
\includegraphics[width=0.74\columnwidth]{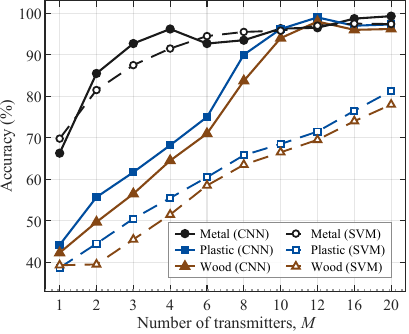}
\caption{Classification accuracy versus $M$ for the three materials under Fibonacci placement at $20$\,dB reference SNR. Solid lines show the CNN and dashed lines the RBF-SVM. The noise floor is calibrated once on metal and held fixed, so lossy materials also experience genuinely lower received SNR. Means over three seeds.}
\label{fig:D1}
\end{figure}

Fig.~\ref{fig:D1} reports accuracy for the three materials. Because $\Gamma$ enters \eqref{eq:channel} once per interaction, material changes the \emph{mixture} of mechanisms reaching the receiver, not merely the received power. For wood the trihedral triple bounce carries $|\Gamma|^{3}$ against $|\Gamma|$ for single bounces, suppressing it by $20\log_{10}|\Gamma|^{2}=28.9$\,dB. With the metal noise floor held fixed, the effective per-bin SNR is $19.9$\,dB for metal, $5.1$\,dB for plastic and $3.6$\,dB for wood. These follow $P/P_{\mathrm{metal}}=(1-f)|\Gamma|^{2}+f|\Gamma|^{6}$, giving $f=0.363$ (plastic) and $0.364$ (wood)---an agreement a single-coefficient $|\Gamma|^{2}$ model cannot reproduce.

Lossy materials therefore need more transmitters, and at low $M$ the requirement is severe. With a single transmitter the CNN reaches $66.3\%$ on metal but only $44.3\%$ on plastic and $42.0\%$ on wood, barely above the $33.3\%$ chance level. Metal passes $90\%$ by $M=3$ while plastic and wood both need $M=10$, an ordering that follows $|\Gamma|$ and is stable across the sweep. By $M=12$ all three converge to $96$--$99\%$, so angular diversity does compensate for material loss, though it takes three to four times as many transmitters to do so.

The classifiers also diverge with material. On metal the SVM tracks the CNN closely and the two end within four points at $M=20$, which confirms that physics-derived features suffice when the signal is strong and well structured. On plastic and wood the SVM curves are still rising at $M=20$ while the CNN curves have saturated, and the gap widens with each additional transmitter rather than closing, reaching $15.3$ points on plastic and $18.0$ points on wood. The hand-crafted features are summary statistics of the magnitude spectrum, and estimating a log-log slope from a spectrum at $4$\,dB SNR is an ill-conditioned regression that degrades quickly as reflectivity falls. The CNN is not restricted to that summary, since it retains the full complex tensor and can exploit relative phase across links, which still carries shape information where a slope estimate has become unreliable. This is the second regime in which the learned model wins clearly, subject to the caveat of Section~\ref{sec:baseline} that part of the gap may follow from phase access rather than from representational capacity.

\subsection{Multi-Object Scenes and Multi-Label Detection}
\label{sec:resC}

Real rooms rarely contain a single isolated target, so we next evaluate scenes holding up to three objects at distinct positions, each with independently drawn azimuth, height scale, position jitter and material. Their returns superpose coherently at the AP, and identification is posed as three independent binary presence tests sharing one input tensor, which is the natural formulation when the number of objects is unknown.

\begin{figure}[t]
\centering
\includegraphics[width=0.74\columnwidth]{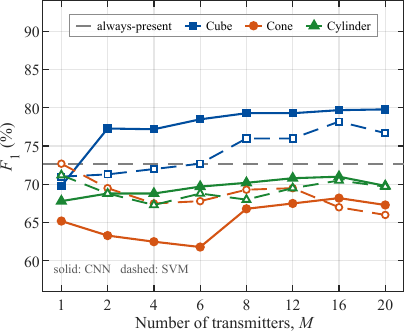}
\caption{Per-shape presence-detection $F_1$ in the three-slot multi-object
scene with mixed materials. Solid lines show the CNN and dashed lines the
RBF-SVM. The horizontal dashed line is the always-present baseline of
$F_1=72.7\%$ at $4/7$ prevalence.}
\label{fig:B1}
\end{figure}

Each shape appears in four of the seven non-empty subsets of the vocabulary, giving a prevalence of $4/7$ and a trivial always-present detector an $F_1$ of $72.7\%$. Fig.~\ref{fig:B1} shows that only the cube detector exceeds this line, rising to a plateau near $80\%$, which is consistent with its angularly sparse signature remaining detectable inside a superposition. The cone and cylinder detectors stay at or below the trivial baseline throughout. More consequentially, all curves saturate by $M\approx4$ and are flat thereafter, and exact-match scene accuracy likewise settles near $0.32$ against a chance value of $0.14$.

Angular diversity, the principal lever in the single-object problem, therefore buys almost nothing once several objects are present, and the limitation has moved to superposition interference. Two properties of the model explain why. Per-sample normalization normalizes the sum, so the strongest scatterer sets the scale and pushes weaker objects toward the effective noise floor. The mechanism decomposition is additive across objects, so the spectral hierarchy reports which mechanisms are present but not how many objects produced them. Additional transmitters supply further views of an already ambiguous superposition without separating it, which makes multi-object identification interference-limited rather than diversity-limited. Progress requires separating the constituent returns before classification rather than adding transmitters.

\subsection{Occlusion and NLoS Robustness}
\label{sec:resD}

At terahertz frequencies indoor propagation is quasi-optical, so partial blockage of the target by furniture or by other users is common. We model occlusion as a fraction $\kappa$ of the target surface hidden from both the illumination and the scattered path, with the occluding half-space fixed in the object frame and azimuth drawn at random, so the shadow direction relative to the transmitters is uniformly distributed. We report two training conditions. In the matched condition the classifier is trained at the same $\kappa$ at which it is tested, which assumes the degree of blockage is known. In the mismatched condition it is trained only on unoccluded targets, which is the deployment-relevant case.

\begin{table}[t]
\caption{Accuracy (\%) under partial occlusion. Metal target, Fibonacci
placement, $20$\,dB SNR, means over three seeds. The first five columns are
matched training. The final column is a classifier trained at $\kappa=0$ and
tested at $M=20$.}
\label{tab:C1}
\centering\scriptsize
\setlength{\tabcolsep}{4pt}
\begin{tabular}{@{}lccccc|c@{}}
\toprule
 & \multicolumn{5}{c|}{matched training} & clean \\
$\kappa$ & $M{=}1$ & $2$ & $4$ & $8$ & $20$ & $M{=}20$ \\
\midrule
0.00 & 72.7 & 84.3 & 94.2 & 91.4 & 98.4 & 98.4 \\
0.15 & 69.9 & 77.2 & 77.9 & 77.9 & 84.1 & 82.9 \\
0.30 & 55.2 & 72.7 & 68.0 & 73.6 & 79.7 & 70.9 \\
0.45 & 54.2 & 68.6 & 69.2 & 74.1 & 74.2 & 67.3 \\
0.60 & 51.2 & 69.2 & 69.2 & 77.0 & 74.8 & 63.6 \\
0.75 & 49.7 & 66.6 & 66.5 & 71.1 & 77.8 & 62.8 \\
\bottomrule
\end{tabular}
\end{table}

Table~\ref{tab:C1} shows a pronounced cliff followed by a plateau. Under matched training the first $15\%$ of occlusion at $M=20$ costs $14$ percentage points, yet hiding a further $60\%$ costs only six more, and at $\kappa=0.75$ accuracy remains near $78\%$. The two stages reflect information sources that degrade on entirely different scales. The sharp initial drop occurs because identification at full aperture relies partly on the coherent interference pattern formed by returns from spatially separated facets, and hiding any part of the surface disrupts it. The plateau reflects what occlusion cannot remove, since a partially occluded cube still presents flat faces meeting at right angles and a partially occluded cylinder still presents a singly curved wall. The classifier continues to read the dominant mechanism type from whatever portion is exposed, and that categorical evidence sustains accuracy well above chance even when most of the object is hidden.

The mismatched column shows that this robustness does not depend on having seen occlusion during training. A classifier trained only on unoccluded targets still reaches $70.9\%$ at $\kappa=0.30$ and $62.8\%$ at $\kappa=0.75$, against a chance level of $33.3\%$. Matched training is worth roughly $9$ to $15$ points across the blocked range, which quantifies the value of knowing the blockage level in advance without making it a precondition for the system to work.

\section{Conclusion}

We have developed a multistatic THz-ISAC framework that identifies indoor object shapes from communication-induced echoes without dedicated sensing hardware, evaluated on orientations held out from training so that the reported accuracies measure generalization rather than closed-set recall.

The PO spectral hierarchy is degenerate in the sense that six physically distinct mechanisms map onto only three frequency exponents, so a spectral slope alone cannot identify surface composition and multistatic angular diversity becomes a requirement for identifiability. The empirical results follow this logic. The cube stays well separated because its discriminant is the angular sparsity of its specular lobes, while the cone and cylinder need several bistatic views to separate tip from curved-surface returns and account for essentially the entire residual error. The CNN outperforms the physics-informed RBF-SVM most clearly under irregular placement, where it gains $8.3$ points, and on lossy materials, where it gains $15$ to $18$ points at an effective SNR below $6$\,dB. Under quasi-uniform placement on metal the two agree closely. Since uncontrolled placement and non-metallic objects are both the norm indoors the advantage is practically relevant, but part of the material gap reflects the access the CNN has to complex phase that the SVM never sees, and separating those two effects remains open.

Under partial occlusion the framework degrades gracefully, and the robustness holds even for a classifier that has never seen an occluded target. Multi-object scenes behave differently, being interference-limited rather than diversity-limited, so separating the constituent returns rather than adding angular diversity is the next problem to solve.

Two limits should be stated plainly. All results derive from a single forward model that has not been checked against full-wave simulation, ray tracing or measurement, and the framework assumes direct-path suppression, multistatic phase coherence and coarse object localization. Surfaces are also treated as optically smooth, whereas at $\lambda=1$\,mm a roughness factor would attenuate the coherent return \cite{swift2006} most strongly for the lossy materials that are already hardest here. Priority future work is a magnitude-only CNN ablation to isolate the phase contribution, a bandwidth sweep as the direct test of the degeneracy thesis, experimental validation at $300$\,GHz, and extension to shapes that share a dominant scattering mechanism.


\end{document}